\documentclass[copyright]{eptcs}
\providecommand{\event}{T. Neary, D. Woods, A.K. Seda and N. Murphy (Eds.):\\ The Complexity of Simple Programs 2008.}
\providecommand{\volume}{1}
\providecommand{\anno}{2009}
\providecommand{\firstpage}{3}
\usepackage{breakurl}

\usepackage{aeguill}
\usepackage{amsmath}
\usepackage{amsfonts}
\usepackage{amsthm}
\usepackage{multirow} 

\newtheorem{remark}{Remark}
\newtheorem{definition}{Definition}
\newtheorem{example}{Example}
\newtheorem{theorem}{Theorem}
\newtheorem{proposition}{Proposition}
\newtheorem{corollary}{Corollary}
\newtheorem{question}{Question}

\newcommand\vx{{x}}

\newcommand\vy{{y}}
\newcommand\vz{{z}}

\newcommand\motnouv[1]{\emph{#1}}
\title{Playing With Population Protocols\footnote{This work and all authors were partly supported by ANR Project SOGEA and by ANR Project SHAMAN, Xavier Koegler was partly supported by COST Action 295 DYNAMO and ANR Project ALADDIN}}

\author{Olivier Bournez
    \email{Olivier.Bournez@lix.polytechnique.fr}
    \institute{Ecole Polytechnique \& Laboratoire d'Informatique (LIX),\\ 91128 Palaiseau Cedex, France}
\and
    J\'er\'emie Chalopin
    \email{Jeremie.Chalopin@lif.univ-mrs.fr}
    \institute{CNRS \& Laboratoire d'Informatique Fondamentale de Marseille, CNRS \& Aix-Marseille Universit{\'e},\\ 39 rue Joliot Curie, 13453 Marseille Cedex 13, France}
\and
    Johanne Cohen
    \email{Johanne.Cohen@prism.uvsq.fr}
    \institute{{CNRS \& PRiSM},\\ 45 Avenue des Etats Unis, 78000 Versailles, France}
\and
    Xavier Koegler
    \email{Xavier.Koegler@liafa.jussieu.fr}
    \institute{\'{E}cole Normale Supérieure \& Université Paris Diderot - Paris 7,\\ Case 7014, 75205 Paris Cedex 13, France}
}
\def\titlerunning{Playing With Population Protocols}
\def\authorrunning{O. Bournez, J. Chalopin, J. Cohen, X. Koegler}

\begin{document}

\maketitle

\begin{abstract}
  Population protocols have been introduced as a model of sensor
  networks consisting of very limited mobile agents with no control
  over their own movement: A collection of anonymous agents, modeled
  by finite automata, interact in pairs according to some rules.

Predicates on the initial configurations that
  can be computed by such protocols have been characterized under
  several hypotheses.

  We discuss here whether and when the rules of interactions between agents
  can be seen as  a game from game theory. We do so by discussing several
  basic protocols.
 
\end{abstract}

\section{Introduction}

The computational power of networks of anonymous resource-limited
mobile agents has been investigated in several recent papers. 

In
particular, Angluin et al.\ proposed in \cite{AspnesADFP2004} a new model of distributed 
computations. In this model, called \motnouv{population
protocols}, finitely many finite-state agents interact in pairs
chosen by an adversary. Each interaction has the effect of updating
the state of the two agents according to a joint transition function.

A protocol is said to \motnouv{(stably) compute} a predicate on the initial states
of the agents if, in any fair execution, after finitely many
interactions, all agents reach a common output that corresponds to the
value of the predicate.

The model was originally proposed to model computations realized by
sensor networks in which passive agents are carried along by other
entities. The canonical example of \cite{AspnesADFP2004} corresponds to
sensors attached to a flock of birds and that must be programmed to
check some global properties, like determining whether more than 5\%
of the population has elevated temperature. Motivating scenarios also
include models of the propagation of trust 
\cite{diamadi2001sgs}.

Much of the work so far on population protocols has concentrated on
characterizing which predicates on the initial states can be computed
in different variants of the model and under various assumptions. In
particular, the  predicates computable by the unrestricted 
population protocols from \cite{AspnesADFP2004} have been characterized
as being precisely the semi-linear predicates, that is to say those
predicates on counts of input agents definable in first-order
Presburger arithmetic \cite{presburger:uvk}. Semilinearity was shown
to be sufficient in \cite{AspnesADFP2004} and necessary in
\cite{AngluinAE2006semilinear}.

Variants considered so far include restriction to one-way
communications, restriction to particular interaction graphs, to
random interactions, with possibly various kind of failures of
agents. Solutions to classical problems of distributed algorithmics
have also been considered in this model. Refer to survey
\cite{PopProtocolsEATCS} for a complete discussion.

The population protocol model shares many features with other models
already considered in the literature. In particular, models of pairwise
interactions have been used to study the propagation of diseases \cite{Heth00}, or rumors
\cite{dk65}. In chemistry the chemical master equation has been
justified using (stochastic) pairwise interactions between the finitely
many molecules present \cite{Murray-VolI,gillespie1992rdc}. In
that sense, the model of population protocols may be considered as fundamental in
several fields of study.
 
Pairwise interactions between finite-state agents are sometimes
motivated by the study of the dynamics of particular two-player games from game theory. For
example, paper \cite{Ref9deFMP04} considers the dynamics of the so-called
$PAVLOV$ behaviour in the iterated prisoner lemma. Several results about
the time of convergence of this particular dynamics towards the stable state can be found in
\cite{Ref9deFMP04}, and \cite{FMP04}, for rings, and complete graphs.

The purpose of the following discussion is to better understand
whether and when pairwise interactions, and hence population protocols, can be considered as the result
of a game. 
We want to understand if restricting to rules that come
from a (symmetric) game is a limitation, and in particular whether restricting to
rules that can be termed $PAVLOV$ in the spirit of \cite{Ref9deFMP04}
is a limitation.
We do so by giving solutions to several basic problems
using rules of interactions associated to a symmetric game.  
As such protocols must also be symmetric,  we are also discussing whether restricting to symmetric
rules in population protocols is a limitation.
 
In Section \ref{section:pp}, we briefly recall population
protocols. In Section \ref{section:gametheory}, we recall some basics
from game theory. In Section \ref{sec:gamepp}, we discuss how a game
can be turned into a dynamics, and introduce the notion of \motnouv{Pavlovian}
population protocol. In Section \ref{sec:results} we prove that any
symmetric deterministic 2-states population protocol is Pavlovian, and that the problem of computing the OR, AND, as well as
the leader election and majority problem admit Pavlovian
solutions. We then discuss our results in Section
\ref{sec:conclusion}.  
  
\section{Population Protocols}
\label{section:pp}

A protocol is given by $(Q,\Sigma,\iota,\omega,\delta)$ with the
following components. $Q$ is a finite set of \motnouv{states}.
$\Sigma$ is a finite set of \motnouv{input symbols}.  $\iota: \Sigma
\to Q$ is the initial state mapping, and $\omega: Q \to \{0,1\}$ is
the individual output function. $\delta \subseteq Q^4$ is a joint
transition relation that describes how pairs of agents can
interact. Relation $\delta$ is sometimes described by listing all
possible interactions using the notation $(q_1,q_2) \to (q'_1,q'_2)$,
or even the notation $q_1q_2 \to q'_1 q'_2$, 
for $(q_1,q_2,q'_1,q'_2) \in \delta$ (with the convention that
$(q_1,q_2) \to (q_1,q_2)$ when no rule is specified with $(q_1,q_2)$
in the left-hand side). The protocol is termed \motnouv{deterministic}
if for all pairs $(q_1,q_2)$ there is only one pair $(q'_1,q'_2)$ with
$(q_1,q_2) \to (q'_1,q'_2)$. In that case, we write
$\delta_1(q_1,q_2)$ for the unique $q'_1$ and $\delta_2(q_1,q_2)$ for
the unique $q'_2$.   

Notice that, in general, rules can be non-symmetric: if
$(q_1,q_2) \to (q'_1,q'_2)$, it does not necessarily follow that $(q_2,q_1) \to
(q'_2,q'_1)$.  

Computations of a protocol proceed in the following way. The
computation takes place among $n$ \motnouv{agents}, where $n \ge 2$. A
\motnouv{configuration} of the system can be described by a vector of
all the agents' states. The state of each agent is an element of $Q$. Because agents
with the same states are indistinguishable, each configuration can be
summarized as an unordered multiset of states, and hence of elements
of $Q$.

Each agent is given initially some input value from $\Sigma$: Each agent's initial
state is determined by applying $\iota$ to its input value. This
determines the initial configuration of the population.

An execution of a protocol proceeds from the initial configuration by
interactions between pairs of agents. Suppose that two agents in state
$q_1$ and $q_2$ meet and have an interaction. They can change into
state $q'_1$ and $q'_2$ if $(q_1,q_2,q'_1,q'_2)$ is in the transition
relation $\delta$.  If $C$ and $C'$ are two configurations, we write
$C \to C'$ if $C'$ can be obtained from $C$ by a single interaction of
two agents: this means that $C$ contains two states $q_1$ and $q_2$ and $C'$ is
obtained by replacing $q_1$ and $q_2$ by $q'_1$ and $q'_2$ in $C$,
where $(q_1,q_2,q'_1,q'_2) \in \delta$. An \motnouv{execution} of the
protocol is an infinite sequence of configurations
$C_0,C_1,C_2,\cdots$, where $C_0$ is an initial configuration and $C_i
\to C_{i+1}$ for all $i\ge0$. An execution is \motnouv{fair} if for
all configurations $C$ that appear infinitely often in the execution,
if $C \to C'$ for some configuration $C'$, then $C'$ appears
infinitely often in the execution.

At any point during an execution, each agent's state determines its
output at that time. If the agent is in state $q$, its output value is
$\omega(q)$. The configuration output is $0$ (respectively $1$) if all
the individual outputs are $0$ (respectively $1$). If the individual
outputs are mixed $0$s and $1s$ then the output of the configuration
is undefined. 

Let $p$ be a predicate over multisets of elements of
$\Sigma$. Predicate $p$ can be considered as a function whose range is
$\{0,1\}$ and whose domain is the collection of these multisets. The predicate is said to be computed by the protocol if,  for every  multiset $I$, and
every fair execution that starts from the initial configuration
corresponding to $I$, the output value of every agent eventually
stabilizes to $p(I)$.

The following was proved in
\cite{AspnesADFP2004,AngluinAE2006semilinear}

\begin{theorem}[\cite{AspnesADFP2004,AngluinAE2006semilinear}] A
  predicate is computable in the population protocol model if and only
  if it is semilinear.
\end{theorem}

Recall that semilinear sets are known to correspond to predicates on
counts of input agents definable in first-order Presburger arithmetic
\cite{presburger:uvk}.

\section{Game Theory}
\label{section:gametheory}

We now recall the simplest concepts from Game Theory. We focus on
non-cooperative games, with complete information, in extensive form.

The simplest game is made up of two players, called $I$ and $II$, with a
finite set of options, called \emph{pure strategies}, $Strat(I)$ and
$Strat(II)$. Denote by $A_{i,j}$ (respectively: $B_{i,j}$) the score
for player $I$
(resp. $II$) when $I$ uses strategy $i \in Strat(I)$ and $II$ uses strategy
$j \in Strat(II)$.
 
The scores are given by $n \times m$ matrices $A$ and $B$, where $n$ and
$m$ are the cardinality of $Strat(I)$ and $Strat(II)$. The game is
termed \emph{symmetric} if $A$ is the transpose of $B$: this implies
that $n=m$, and we can assume without loss of generality that
$Strat(I)=Strat(II)$.

\begin{example}[Prisoner's dilemma]
The case where $A$ and $B$ are the following matrices 

$$A = \left(
\begin{array}{ll}
R & S \\
T & P \\
\end{array}
\right)
, 
B= 
\left(
\begin{array}{ll}
R & T \\
S & P \\
\end{array}
\right)
$$
with $T>R>P>S$ and $2R >T+S$, is called the \emph{prisoner's dilemma}. We denote by $C$ (for
cooperation) the first pure strategy, and by $D$ (for defection) the
second pure strategy of each player.

As the game is symmetric, matrix $A$ and $B$ can also be denoted by:

\begin{center}
\begin{tabular}{llcr} 
  &   & \multicolumn{2}{c}{Opponent}  \\
  &  &  { C} &{ D } \\ \cline{3-4}
\multirow{2}{*}{Player} & C &  \multicolumn{1}{|c}{$R$} &\multicolumn{1}{r|}{$S$}\\
                        & D &  \multicolumn{1}{|c}{$T$} &\multicolumn{1}{r|}{$P$}\\ \cline{3-4}
\end{tabular}
\end{center}
\end{example}





A strategy $x \in Strat(I)$ is said to be a best response to strategy
$y \in Strat(II)$, denoted by $x \in BR(y)$ if
\begin{equation}
A_{z,y} \le A_{x,y}
\end{equation}
for all strategies $z \in Strat(I)$.


A pair $(x,y)$ is a \emph{(pure) Nash equilibrium} if $x \in
BR(y)$ and $y \in BR(x)$. A pure Nash equilibrium does not always
exist. 


In other words, two strategies $(x,y)$ form a Nash equilibrium if
in that state neither of the players has a unilateral interest to deviate
from it.


\begin{example}
  On the example of the prisoner's dilemma, $BR(\vy)=D$ for all
  $\vy$, and $BR(\vx)=D$ for all $\vx$. So $(D,D)$ is the unique
  Nash equilibrium, and it is pure. In it, each player has score
  $P$. The paradox is that if they had played $(C,C)$ (cooperation)
  they would have had score $R$, that is more. The social optimum
  $(C,C)$, is different from the equilibrium that is reached by
  rational players $(D,D)$, since in any other state, each player
  fears that the adversary plays $C$.
\end{example}

We will also introduce the following definition: Given some strategy $\vx' \in Strat(I)$, a strategy $\vx \in Strat(I)$ is said to be a best response to strategy
$\vy \in Strat(II)$ among those different from $\vx'$, denoted by $\vx
\in BR_{\neq \vx'}(\vy)$ if
\begin{equation}
A_{z,y} \le A_{x,y}
\end{equation}
for all strategy $\vz \in Strat(I), \vz \neq \vx'$. 

Of course, the role of $II$ and $I$ can be inverted in the previous definition.


There are two main approaches to discussing dynamics of games. The first
consists in repeating games. The second in using models from
evolutionary game theory. Refer to \cite{Evolutionary1,LivreWeibull}
for a presentation of this latter approach.


\paragraph{Repeating Games.}

Repeating $k$ times a game, is equivalent to extending the space of
choices into $Strat(I)^k$ and $Strat(II)^k$: player $I$ (respectively
$II$) chooses his or her action $\vx(t) \in Strat(I)$, (resp. $\vy(t) \in
Strat(II)$) at time $t$ for $t=1,2,\cdots,k$.  Hence, this is
equivalent to a two-player game with respectively $n^k$ and $m^k$
choices for players.

To avoid confusion, we will call
\emph{actions} the choices $\vx(t),\vy(t)$ of each player at a given time,
and \emph{strategies} the sequences $X=\vx(1),\cdots,\vx(k)$ and
$Y=\vy(1),\cdots,\vy(k)$, that is to say the strategies for the global
game.

If the game is repeated an infinite number of times, a strategy
becomes a function from integers to the set of actions, and the game
is still equivalent to a two-player game\footnote{but whose matrices
  are infinite.}.

\paragraph{Behaviours.}

In practice, player $I$ (respectively $II$) has to solve the following
problem at each time $t$: given the history of the game up to now,
that is to say $$X_{t-1}=\vx(1),\cdots,\vx(t-1)$$ and
$$Y_{t-1}=\vy(1),\cdots,\vy(t-1)$$ what should I play at time $t$? In
other words, how to choose $\vx(t) \in Strat(I)$?  (resp. $\vy(t) \in
Strat(II)$?)

Is is natural to suppose that this is
given by some behaviour rules: $$\vx(t)=f(X_{t-1},Y_{t-1}),$$
$$\vy(t)=g(X_{t-1},Y_{t-1})$$ for some particular functions $f$ and $g$.

\paragraph{The Specific Case of the Prisoner's Lemma.} 
The question of the best behaviour rule to use for the prisoner lemma
gave birth to an important literature. In particular, after the book
\cite{axelrod:1984:ec}, that describes the results of tournaments of
behaviour rules for the iterated prisoner lemma, and that argues that
there exists a best behaviour rule called $TIT-FOR-TAT$.
This  consists in cooperating at the first step,
and then do the same thing as the adversary at subsequent times. %

A lot of other behaviours, most of them with very
picturesque names have been proposed and studied: see for example
\cite{axelrod:1984:ec}, \cite{TheseBeaufils}, \cite{DEALabbani}.

Among possible behaviours is $PAVLOV$: in the iterated
prisoner lemma, a player cooperates if and only if both players opted
for the same alternative in the previous move. This name 
\cite{kraines1988psd,nowak1993sws,axelrod:1984:ec} stems from the fact that this strategy embodies
an almost reflex-like response to the payoff: it repeats its former
move if it was rewarded by $R$ or $T$ points, but switches behaviour if
it was punished by receiving only $P$ or $S$ points. Refer to
\cite{nowak1993sws} for some study of this strategy in the spirit of
Axelrod's tournaments.

The $PAVLOV$ behaviour can also be termed \textit{WIN-STAY, LOSE-SHIFT} as
if the play on the previous round resulted in a success, then the
agent plays the same strategy on the next round. Alternatively, if the
play resulted in a failure the agent switches to another action
\cite{nowak1993sws,axelrod:1984:ec}.

\paragraph{Going From $2$ Players to $N$ Players.} $PAVLOV$ behaviour is Markovian: a behaviour $f$ is \emph{Markovian}, if
$f(X_{t-1},Y_{t-1})$ depends only on $\vx(t-1)$ and $\vy(t-1)$.

From such a behaviour, it is easy to obtain a distributed
dynamic. For example, let's follow \cite{Ref9deFMP04}, for the
prisoner's dilemma.

Suppose that we have a connected graph $G=(V,E)$, with $N$
vertices. The vertices correspond to players. An instantaneous
configuration of the system is given by an element of $\{C,D\}^N$,
that is to say by the state $C$ or $D$ of each vertex. Hence, there
are $2^N$ configurations.

At each time $t$, one chooses randomly and uniformly one edge $(i,j)$
of the graph. At this moment, players $i$ and $j$ play the prisoner
dilemma with the $PAVLOV$ behaviour. It is easy to see that this
corresponds to executing the following rules:

\begin{equation} \label{eq:pavlov}
\left\{
\begin{array}{lll}
CC &\to& CC \\
CD &\to& DD \\
DC &\to& DD \\
DD &\to& CC. \\
\end{array}
\right.
\end{equation}

What is the final state reached by the system?  The underlying model is a very large Markov chain with $2^N$ states. The
state $E^*=\{C\}^N$ is absorbing. If the graph $G$ does not have any
isolated vertex, this is the unique absorbing state, and there exists
a sequence of transformations that transforms any state $E$ into this
state $E^*$.
As a consequence, from well-known classical results in Markov chain theory, whatever  the initial configuration is, with
probability $1$, the system will eventually be in state $E^*$ 
\cite{Markov}. The system is \emph{self-stabilizing}.

Several results about the time of convergence  towards this stable state
can be found in \cite{Ref9deFMP04}, and \cite{FMP04}, for  rings,
and  complete graphs.

What is interesting in this example is that it shows how to go from a 
game, and a behaviour to a distributed dynamics on a
graph, and in particular to a population protocol when the graph is
the complete graph.  

\section{From Games To Population Protocols}
\label{sec:gamepp}

In the spirit of the previous discussion, to any symmetric game, we can
associate a population protocol as follows.

\begin{definition}[Associating a Protocol to a Game]
  Assume a symmetric two-player game is given. Let $\Delta$ be
  some threshold. 
  
  The  protocol associated to the game  is a population
  protocol whose set of states is $Q$, where  $Q=Strat(I)=Strat(II)$
  is the set of strategies of the game, and whose transition rules
  $\delta$ are
  given as follows:
$$(q_1,q_2,q'_1,q'_2) \in \delta$$ where
  \begin{itemize}
  \item $q'_1=q_1$ when $M_{q_1,q_2}\ge\Delta$
  \item $q'_1 \in BR_{\neq q_1}(q_2)$ when $M_{q_1,q_2} < \Delta$
  \end{itemize}
and
   \begin{itemize}
  \item $q'_2=q_2$ when $M_{q_2,q_1} \ge \Delta$
  \item $q'_2 \in BR_{\neq q_2}(q_1)$ when $M_{q_2,q_1} < \Delta$,
  \end{itemize}
where $M$ is the matrix of the game.
\end{definition}

\begin{definition}[Pavlovian Population Protocol]
A population protocol is \motnouv{Pavlovian} if it can be obtained
from a game as above.
\end{definition}

\begin{remark}
Clearly a Pavlovian population protocol must be \motnouv{symmetric}:
indeed, whenever
$(q_1,q_2,q'_1,q'_2) \in \delta$, one has $(q_2,q_1,q'_2,q'_1) \in
\delta$.
\end{remark}

\section{Some Specific Pavlovian Protocols} 
\label{sec:results}


We now discuss whether assuming protocols Pavlovian is a restriction.

We start by an easy consideration.

\begin{theorem} Any symmetric deterministic $2$-states population protocol is Pavlovian.
\end{theorem}

\begin{proof}
  Consider a deterministic symmetric $2$-states population protocol. Note $Q=\{+,-\}$ its set
  of states. Its transition function can be written as follows:
 
\begin{equation}
\left\{
\begin{array}{lll}
++ & \to & \alpha_{++} \alpha_{++}  \\
+- & \to &  \alpha_{+-} \alpha_{-+}\\
-+ & \to & \alpha_{-+} \alpha_{+-}\\
-- & \to & \alpha_{--}  \alpha_{--}\\
\end{array}
\right.\ 
\end{equation}
for some  $\alpha_{++},\alpha_{+-},\alpha_{-+},\alpha_{--}$.

This corresponds to the symmetric game given by the following pay-off matrix $M$  
\begin{center}
\begin{tabular}{llcc} 
  &   & \multicolumn{2}{c}{Opponent}  \\
  &  &  \multicolumn{1}{c}{ \textsc{+}}     & { \textsc{-} }  \rule[-7pt]{0pt}{20pt}\\ \cline{3-4}
 \multirow{2}{*}{Player} & \textsc{+} &  \multicolumn{1}{|c}{$\beta_{++}$}  &\multicolumn{1}{c|}{$\beta_{+-}$}\rule[-7pt]{0pt}{20pt}\\
                        & \textsc{-}  &  \multicolumn{1}{|c}{$\beta_{-+}$}  &\multicolumn{1}{c|}{$\beta_{--}$}\rule[-7pt]{0pt}{20pt}\\ \cline{3-4}
\end{tabular}
\end{center}
taking threshold $\Delta=1$, where for all $q_1,q_2 \in \{+,-\}$,
\begin{itemize}
\item  $\beta_{q_1q_2}=2$ if  $\alpha_{q_1 q_2}=q_1$,
\item $\beta_{q_1q_2}=0$  otherwise.
\end{itemize}


 
\end{proof}

Unfortunately, not all rules correspond to a game.

\begin{proposition}
Some symmetric population protocols are not Pavlovian.
\end{proposition}

\begin{proof}
Consider for example a deterministic $3$-states population protocol
with set of states $Q=\{q_0,q_1,q_2\}$ and a joint transition function
$\delta$ such that $\delta_1(q_0,q_0)=q_{1}$,
$\delta_1(q_1,q_0)=q_{2}$ , $\delta_1(q_2,q_0)=q_{0}$.

Assume by contradiction that there exists a $2$-player game
corresponding to this $3$-states population protocol. Consider its
payoff matrix $M$. Let $M(q_0,q_0)=\beta_{0}$, $M(q_1,q_0)=\beta_{1}$
, $M(q_2,q_0)=\beta_{2}$. We must have $\beta_0 \ge \Delta, \beta_1
\ge \Delta$ since all agents that interact with
an agent in state $q_0$ must change their state. Now, since $q_0$
changes to $q_1$, $q_1$ must be a strictly better response to $q_0$ than
$q_2$: hence, we must have $\beta_1 > \beta_2$.  In a similar way,
since $q_1$ changes to $q_2$, we must have $\beta_2 > \beta_0$ , and
since $q_2$ changes to $q_0$, we must have $\beta_0 > \beta_1$. From
$\beta_1 > \beta_2 > \beta_0$ we reach a contradiction.
\end{proof}

This indeed motivates the following study, where we discuss which problems
admit a Pavlovian solution.

\subsection{Basic Protocols}

\begin{proposition} There is a Pavlovian protocol that computes the
  logical $OR$ (resp. $AND$) of input bits.
\end{proposition}

\begin{proof} 
Consider the following protocol to compute $OR$, 

\begin{equation}
\left\{
\begin{array}{lll}
01 & \to & 11 \\
10 & \to & 11  \\
00 & \to & 00 \\
11 & \to & 11\\
 \end{array}
\right.
\end{equation}

and the following protocol to compute $AND$, 

\begin{equation}
\left\{
\begin{array}{lll}
01 & \to & 00 \\
10 & \to & 00  \\
00 & \to & 00 \\
11 & \to & 11\\
 \end{array}
\right.
\end{equation}

Since they are both deterministic 2-states population protocols,
they are
Pavlovian. 

\end{proof}

\begin{remark}
Notice that $OR$  (respectively $AND$) protocol corresponds to the predicates on counts of
input agents $n_0 \ge 1$ (resp. $n_1=0$) where $n_0$, $n_1$ are the number of input agents in
state $0$ and $1$ respectively.
\end{remark}

\begin{remark}
All previous protocols are ``naturally broadcasting'' i.e., eventually all agents
agree on some (the correct) value. With previous definitions (which are the
classical ones for population protocols), the  following protocol
does not compute the $XOR$ or input bits, or equivalently does not
compute predicate $n_1 \equiv
1~ (mod~2)$.  \begin{equation}
\left\{
\begin{array}{lll}
01 & \to & 01 \\
10 & \to & 10  \\
00 & \to & 00 \\
11 & \to & 00\\
 \end{array}
\right.
\end{equation}

Indeed, the answer is not eventually known
by all the agents. It computes the $XOR$ in a weaker form i.e., eventually, all
agents will be in state $0$, if the $XOR$ of input bits is $0$, or
eventually only one agent will be in state $1$, if the $XOR$ of input
bits is $1$.
\end{remark}

\subsection{Leader Election}

The classical solution \cite{AspnesADFP2004} to the leader election problem (starting
from a configuration with $\ge 1$ leaders, eventually exactly one leader
survives) is the following:

\begin{equation} 
\left\{
\begin{array}{lll}
LL & \to & LN \\
LN & \to & LN  \\
NL & \to & NL \\
NN & \to & NN\\
 \end{array}
\right.
\end{equation}

Unfortunately, this protocol is  non-symmetric, and hence
non-Pavlovian.

\begin{remark} Actually, the problem is
with the first rule, since one wants two leaders to become only
one. If the two leaders are identical, this is clearly problematic with symmetric rules.
\end{remark}

However, the leader election problem can actually be solved by a Pavlovian
protocol, at the price of a less trivial protocol.




\begin{proposition}
The following Pavlovian protocol solves the leader election problem,
as soon as the population is of size $\ge 3$.
\begin{equation} \label{ode:dynamic}
\left\{
\begin{array}{lll}
L_1L_2 & \to & L_1N \\
L_1N& \to & NL_2\\
L_2N & \to & NL_1 \\
NN & \to & NN\\
L_2L_1 & \to & N L_1 \\
N L_1 & \to &  L_2 N \\
N L_2 & \to &  L_1  N\\
L_1 L_1 & \to &L_2 L_2 \\
L_2 L_2 & \to &L_1 L_1 \\
 \end{array}
\right.
\end{equation}
\end{proposition}

\begin{proof}
Indeed, starting from a configuration containing not only $N$s,
eventually after some time configurations will have exactly one
leader, that is one agent in state $L_1$ or $L_2$.

Indeed, the first rule and the fifth rule decrease strictly the
number of leaders whenever there are more than two leaders. Now the
other rules, preserve the number of leaders, and are made such that an $L_1$ can always be transformed
into an $L_2$ and vice-versa, and hence are made such that a configuration where
first or fifth rule applies can always be reached whenever there are
more than two leaders. The fact that it solves the leader election
problem then 
follows from the hypothesis of fairness in the definition of
computations.

This is a Pavlovian protocol, since it corresponds to the following
payoff matrix, with threshold $\Delta=4$

\begin{center}
\begin{tabular}{llccc} 
  &   & \multicolumn{3}{c}{Opponent}  \\
  &  &  { \textsc{$L_1$}}     & { \textsc{$L_2$}  } & {\textsc{N}}  
\rule[-7pt]{0pt}{20pt}\\ \cline{3-5}
 \multirow{3}{*}{Player} & \textsc{$L_1$} &  \multicolumn{1}{|c}{$1$}
 & \multicolumn{1}{c}{$4$}
 &\multicolumn{1}{c|}{$1$}\rule[-7pt]{0pt}{20pt}\\

 & \textsc{$L_2$} &  \multicolumn{1}{|c}{$3$}
 & \multicolumn{1}{c}{$1$}
 &\multicolumn{1}{c|}{$1$}\rule[-7pt]{0pt}{20pt}\\

 & \textsc{$N$} &  \multicolumn{1}{|c}{$2$}
 & \multicolumn{1}{c}{$1$}
 &\multicolumn{1}{c|}{$4$}\rule[-7pt]{0pt}{20pt}\\
\cline{3-5}

\end{tabular}
\end{center}
\end{proof}

\subsection{Majority}

\begin{proposition}
The majority problem (given some population of $0$s and $1$s, determine
whether there are more $0$s than $1$s) can be solved by a Pavlovian population protocol.
\end{proposition}

If one prefers, the predicate $n_0 \ge n_1$ on counts of input agents can be computed by a Pavlovian
population protocol.

\begin{proof}

  We claim that the following  protocol outputs 1 if there
  are more $0$s than $1$s in the initial configuration and 0
  otherwise, 

\begin{equation} 
\left\{
\begin{array}{lll}
NY & \to & YY \\
YN & \to & YY \\

N0 & \to & Y0 \\
0N & \to & 0Y \\

Y1 & \to & N1 \\
1Y & \to & 1N \\

01 & \to & NY \\
10 & \to & YN \\
 \end{array}
\right.
\end{equation}

taking 
\begin{itemize}
\item $\Sigma = \{0,1\}, Q = \{0,1,Y,N\}$, 
\item $\omega(0) = \omega(Y) = 1$,
\item $\omega(1) = \omega(N) = 0$.
\end{itemize}

In this protocol, the states $Y$ and $N$ are ``neutral'' elements for
our predicate but they should be understood as \emph{Yes} and
\emph{No}. They are the ``answers'' to the question: are there more $0$s
than $1$s.

This protocol is made such that the number of $0$s and $1$s is
preserved except when a $0$ meets a $1$. In that latter case, the two
agents are deleted and transformed into a $Y$ and a $N$. 

If there are initially strictly more $0$s than $1$s, from the fairness
condition, each $1$ will be paired with a $0$ and at some point no
$1$ will left. By fairness and since there is still at least a
$0$, a configuration containing only $0$ and $Y$s will be
reached. Since in such a configuration, no rule can modify the state
of any agent, and since the output is defined and equals to $1$ in
such a configuration, the protocol is correct in this case

By symmetry, one can show that the protocol outputs $0$ if there are
initially strictly more $1$s than $0$s. 

Suppose now that initially, there are exactly the same number of $0$s
and $1$s. By fairness, there exists a step when no more
agents in the state $0$ or $1$ left. Note that at the moment where the last
$0$ is matched with the last $1$, a $Y$ is created. Since this $Y$ can
be ``broadcast'' over the $N$s, in the final configuration all
agents are in the state $Y$ and thus the output is correct.

This protocol is Pavlovian, since it corresponds to the following
payoff matrix with threshold~$2$.

\begin{center}
\begin{tabular}{llcccc} 
  &   & \multicolumn{4}{c}{Opponent}  \\
  &  &  \multicolumn{1}{c}{ ~\textsc{N}~~} & \textsc{Y} & \textsc{0} & { \textsc{1} } \\ \cline{3-6}
 & \textsc{N} &  \multicolumn{1}{|c}{~$3$~~} & $1$  & $1$
  &\multicolumn{1}{c|}{$3$}\\

Player~~~ & \textsc{Y} &  \multicolumn{1}{|c}{~$2$~~} & $3$ & $3$
&\multicolumn{1}{c|}{$1$}\\

&  \textsc{0}   &  \multicolumn{1}{|c}{~$2$~~} & $2$ & $2$
&\multicolumn{1}{c|}{$1$}\\ 

&  \textsc{1}   &  \multicolumn{1}{|c}{~$2$~~} & $2$ &$1$
&\multicolumn{1}{c|}{$2$}\\

\cline{3-6}
\end{tabular}
\end{center}

\end{proof}

\section{Discussions}
\label{sec:conclusion}

We proved that predicates on counts of input agents $n \ge 0$, $n =
0$, $n \ge m$, where $n,m$ are some counts of input agents, can be
computed by some Pavlovian population protocols.

It is clear that the subset of the predicates computable by
Pavlovian population protocols is closed by negation: just switch the
value of the individual output function of a protocol computing a
predicate to get a protocol computing its negation.

However, some work remains to be done to fully characterize which predicates can
be computed by a Pavlovian population protocol. The first steps would be
to understand the following questions.

\begin{question} 
Is $mod~2$, or equivalently the predicate $n  \equiv
1~ (mod~2)$,  computable by a Pavlovian population
  protocol?
\end{question}

\begin{question} 
Is $\ge k$, or equivalently the predicate $n  \ge k$,  for fixed
$k$, computable by a Pavlovian population
  protocol?
\end{question}

Notice that, unlike what happens for general population protocols, 
composing Pavlovian population protocols into a Pavlovian population
protocol is not easy. It is not clear whether Pavlovian
computable predicates are closed by conjunctions: classical
constructions for general population protocols can not be used directly.

As we said, Pavlovian Population protocols are symmetric. We however
know that assuming population protocols symmetric is not a
restriction.

\begin{proposition}
Any  population protocol can be simulated by a symmetric  population
protocol, as soon as the population is of size $\ge 3$.
\end{proposition}

Before proving this proposition, we state the (immediate) main consequence.

\begin{corollary}
A
  predicate is computable by a symmetric population protocol if and only
  if it is semilinear.
\end{corollary}

\begin{proof}[Proof (of proposition):]
To a population protocol $(Q,\Sigma,\iota,\omega,\delta)$, with
$Q=\{q_1,\cdots,q_n\}$ associate population protocol
$(Q \cup Q',\Sigma,\iota,\omega,\delta')$ with
$Q'=\{q'_1,\cdots,q_n'\}$, $\omega(q')=\omega(q)$ for all $q \in Q$, and for all rules
$$qq \to \alpha \beta$$ in $\delta$, the following rules in $\delta'$:
$$
\left\{
\begin{array}{lll}
qq' &\to& \alpha\beta \\
q'q &\to& \beta\alpha \\
qq &\to& q'q' \\
q'q' &\to& qq \\
q \gamma &\to & q' \gamma \\
q' \gamma &\to& q \gamma \\
\gamma q & \to & \gamma q' \\
\gamma q' & \to & \gamma q \\
\end{array}
\right.
$$
for all $\gamma \in Q \cup Q', \gamma \neq q, \gamma \neq q'$,
and for all pairs of rules
$$
\left\{
\begin{array}{lll}
qr &\to& \alpha\beta \\
rq &\to& \delta\epsilon \\
\end{array}
\right.
$$
with $q,r \in Q$, the following rules in $\delta'$:
$$
\left\{
\begin{array}{lll}
qr' &\to& \alpha\beta \\
r'q & \to & \beta \alpha \\
rq' & \to & \delta \epsilon \\
q'r & \to & \epsilon \delta. \\
\end{array}
\right.
$$
The obtained population protocol is clearly symmetric. Now the first
set of rules guarantees that a state in $Q$ can always be converted to
its primed version in $Q'$ and vice-versa. By fairness, whenever a
rule $qq \to \alpha\beta$ (respectively $qr \to \alpha\beta$) can be applied, then the corresponding two first
rules of the first set of rules (resp. of the second set of rules) can
eventually be fired after possibly some conversions of states into their
primed version or vice-versa.
\end{proof}

\bibliographystyle{plain}

\bibliographystyle{eptcs}

\end{document}